\documentclass{elsart}
\usepackage{graphicx,amssymb,bm}
\journal{Physics Letters A}
\begin{document}
\begin{frontmatter}
\title{Two-soliton solution for the derivative nonlinear
Schr\"odinger equation with nonvanishing boundary conditions}
\thanks{This work was supported by the National Natural Science Foundation of China
under Grant No. 10375027 and the SRF for ROCS, SEM of China.}
\author{Xiang-Jun Chen,\corauthref{xjc}}
\corauth[xjc]{Corresponding author.}
\ead{xiangjun-chen@21cn.com}
\author{Hui-Li Wang,}
\author{and Wa Kun Lam}
\address{Department of Physics, Jinan University, Guangzhou 510632, P. R. China}
\begin{abstract}
An explicit two-soliton solution for the derivative nonlinear Schr\"odinger equation
with nonvanishing boundary conditions is derived, demonstrating details of
interactions between two bright solitons, two dark solitons, as well as one bright
soliton and one dark soliton.
Shifts of soliton positions due to collisions are analytically obtained, which are
irrespective of the bright or dark characters of the participating solitons.
\end{abstract}
\end{frontmatter}
The derivative nonlinear Schr\"odinger (DNLS)
equation is an integrable model describing various nonlinear waves such as
nonlinear Alfv\'en waves in space plasma(see, e.g.,
\cite{Rogister71,Mjolhus76,Mio76,Mjolhus89,Mjolhus97,Ruderman02,Ruderman02b}),
sub-picosecond pulses in single mode optical fibers(see, e.g.,
\cite{Agrawal01,Tzoar,Anderson83,Ohkuma87,Doktorov}), and weak nonlinear electromagnetic
waves in ferromagnetic \cite{Nakata91}, antiferromagnetic\cite{Daniel02}, and
dielectric\cite{Nakata93} systems under external magnetic fields.
Both of vanishing boundary conditions (VBC) and nonvanishing boundary conditions (NVBC)
for the DNLS equation are physically significant.
For problems of nonlinear Alfv\'en waves, weak nonlinear electromagnetic
waves in magnetic and dielectric media, waves propagating strictly parallel to the
ambient magnetic fields are modelled by VBC while those oblique waves are modelled by
NVBC. In optical fibers, pulses under bright background waves are
modelled by NVBC.

It has been known that integrable systems admit soliton solutions
which pass through each other in a completely elastic fashion. The only traces of their
interactions
are shifts of positions and phases. In some systems, besides ``pure'' solitons which
keep their forms unchanged in transmissions, there exist bound states of
these pure solitons called breathers which periodically oscillate
(see, e.g., \cite{Lamb}).

Soliton solutions for the DNLS equation with VBC, including one-soliton
solution\cite{Kaup78} and multi-soliton formulas(e.g.,\cite{Nakamura,Huang90}), have
been known. Researches on the DNLS equation with NVBC showed that its
general one-soliton solution (corresponding to a complex discrete spectral parameter)
is a breather which degenerates to a pure bright or dark soliton when the discrete
spectral parameter becomes purely
imaginary\cite{Kawata78,Kawata79,Mjolhus89,Mjolhus97,Chen04}.
In known (1+1)-dimensional one-component integrable systems, the DNLS equation with NVBC is
a rare instance which simultaneously supports pure bright
solitons, pure dark solitons, as well as their bound states.
Collisions between these solitons are thus important topics.
However, like other NVBC problems,
a double-valued function of the spectral parameter inevitably appears in the
inverse scattering transform (IST) for the DNLS equation with NVBC, greatly complicating the
IST. Early IST
for the system performed on Riemann sheets only obtained modulus of a one-soliton
solution\cite{Kawata78} and asymptotic behaviors of the modulus of an
implicit pure two-soliton solution\cite{Kawata79}. Although the phase of the
one-soliton solution was found later, yielding a very complicated
solution\cite{Mjolhus89,Mjolhus97}, it is a too tedious task to get an explicit
multi-soliton solution based on the IST performed on Riemann sheets
\cite{Kawata78,Kawata79}.
A recent multi-soliton formula using B\"acklund transformation\cite{Steudel03} was
also unable to explicitly demonstrate collisions between solitons.

It has been suggested that constructing Riemann sheets for such NVBC problems can
be avoided if one performs the IST on the plane of an appropriate affine
parameter\cite{FaddeevBK}. The technique
was recently applied to the DNLS equation with NVBC, yielding not only a much simpler
one-soliton
solution than those in the literature but also a simple IST for
further researches\cite{Chen04}.
Immediately following Ref. \cite{Chen04}, infinite number of conservation laws was
derived by a simple standard procedure\cite{Chen05} and the evolution of a
rectangular initial pulses in the system was considered, which was shown to be highly
nontrivial and significantly different from all known results of other integrable
systems\cite{Lashkin}.

It should be emphasized that these DNLS solitons might not exist in all of the physical
systems mentioned above when the (1+1)-dimensional DNLS equation is not rigorously
valid. For example, in space plasma, when the DNLS equation was
generalized to a multi-dimensional one\cite{Mjolhus86,Ruderman87},
it was shown that the dark solitons are not stable in presence of transverse
perturbations\cite{Ruderman87}.

In this letter, we only consider pure solitons of the DNLS equation with NVBC.
We modify the IST in
Ref.\cite{Chen04} to the case when all discrete spectral parameters are
purely imaginary and derive an explicit pure two-soliton solution.
The solution consists of two bright solitons, two dark solitons, or one bright
soliton and one dark soliton. Shifts of soliton positions due to collisions between
them are obtained. We find that the collision between a bright soliton and a dark
soliton is similar to usual collisions between two bright solitons or two dark
solitons:
the position of the faster one get a forward shift while the position of the slower one
get a backward shift. We also show that the shifts of soliton positions are
irrespective
of the bright or dark characters of those participating solitons.

We write the DNLS equation as
\begin{equation}\label{DNLSeq}
iu_t+u_{xx}+i(|u|^2u)_x=0,
\end{equation}
where the subscript denotes partial derivative. Its first Lax equation\cite{Kaup78}
is
\begin{equation}\label{Lax1}
\partial_xF=LF,
\end{equation}
with
\begin{equation}
L=-i\lambda^2\sigma_3+\lambda U,
\end{equation}
\begin{equation}
U=\left(\begin{array}{cc}0 & u \cr -\bar{u} & 0\end{array}\right).
\end{equation}
Here $\sigma_i(i=1,2,3)$ are Pauli matrices,
\begin{equation}
\sigma_1=\left(\begin{array}{cc}0 & 1 \\ 1 & 0\end{array}\right),\quad
\sigma_2=\left(\begin{array}{cc}0 & -i \\ i & 0\end{array}\right),\quad
\sigma_3=\left(\begin{array}{cc}1 & 0 \\ 0 & -1\end{array}\right),
\end{equation}
the bar stands for
complex conjugate, and $\lambda$ is the time-independent
spectral parameter.
As there is no phase shift across the DNLS solitons with NVBC\cite{Mjolhus89,Chen04},
the NVBC can be simply written as
\begin{equation}
u\to\rho, \quad \mbox{as} \quad x\to\pm\infty,
\end{equation}
where $\rho$ is real.
Asymptotic solutions of Eq.~(\ref{Lax1}) is
\begin{equation}
E^\pm(x,k)=(I-i\rho k^{-1}\sigma_1)e^{-i\lambda\zeta x\sigma_3},
\quad \mbox{as} \quad x\to\pm\infty,
\end{equation}
where we introduce an affine parameter $k$ to make
$\zeta=(\lambda^2+\rho^2)^{\frac12}$,
a double-valued function of $\lambda$, become single-valued function of $k$, with
\begin{equation}
\lambda=\frac12(k-\rho^2k^{-1}), \quad \zeta=\frac12(k+\rho^2k^{-1}).
\end{equation}
We define Jost solutions,
\begin{equation}
\Psi(x,k)\to E^+(x,k), \quad \mbox{as}\quad
x\to\infty,
\end{equation}
\begin{equation}
\Phi(x,k )\to E^-(x,k), \quad \mbox{as}\quad
x\to-\infty,
\end{equation}
where
\begin{equation}
\Psi(x,k)=\left(\tilde{\psi}(x,k ),\quad \psi(x,k )\right),
\end{equation}
\begin{equation}
\Phi(x,k )=\left(\phi(x,k ),\quad \tilde{\phi}(x,k )\right),
\end{equation}
and the scattering coefficients by
\begin{equation}\label{Jost}
\phi(x,k )=a(k)\tilde{\psi}(x,k)+b(k)\psi(x,k),
\end{equation}
\begin{equation}
\tilde{\phi}(x,k )=\tilde{a}(k)\psi(x,k)-\tilde{b}(k)\tilde{\psi}(x,k).
\end{equation}
$\psi(x,k)$, $\phi(x,k)$ and $a(k)$ are analytic in the first and the third quadrants
of the complex $k$ plane, while $\tilde{\psi}(x,k)$, $\tilde{\phi}(x,k)$ and
$\tilde{a}(k)$ are analytic in the second and the fourth quadrants.

As shown in Ref.\cite{Chen04}, on the plane of the affine parameter $k$,
if $k_{n1}=k_n$ is a simple zero of $a(k)$ in
the first quadrant, then $k_{n2}=-k_n$, $k_{n3}=\rho^2\bar{k}_n^{-1}$, and
$k_{n4}=-\rho^2\bar{k}_n^{-1}$ are also simple zeros. For the case when
all discrete parameters $\lambda_n (n=1,2,\dots,N)$ are purely imaginary, all zeros of
$a(k)$ locate on the circle of radius $\rho$ centered at the origin($\rho$-circle), that is
\begin{equation}
k_n=\rho \exp(i\beta_n), \quad 0<\beta_n<\pi/2, \quad n=1,2,\dots, N.
\end{equation}
Then, $k_{n3}=k_{n1}$, $k_{n4}=k_{n2}$, contributions of $k_{n3}$ and $k_{n4}$
must be dropped from relevant equations obtained in Ref.\cite{Chen04}.
Therefore, for reflectionless potentials, we get
\begin{equation}\label{ak}
a(k)=e^{i\eta}\prod_{n=1}^{N}\frac{k^2-k_n^2}{k^2-\bar{k}_n^2},
\end{equation}
where $\eta=-2\sum_n\beta_n$, the inverse scattering equation,
\begin{equation}\label{IST}\nonumber
\tilde{\psi}(x,k)e^{i\lambda\zeta x}=\left(\begin{array}{cc}e^{-i\eta^+} \cr -i\rho
k^{-1}e^{i\eta^+}\end{array}\right)+2\sum_{n=1}^N\left(\begin{array}{cc}k_n & 0 \cr 0 & k
\end{array}\right)\frac{c_n\psi(x,k_n)}{k^2-k_n^2}e^{i\lambda_n \zeta_nx},
\end{equation}
and the expression for reflectionless potentials (solitons),
\begin{equation}\label{u}
u(x)=\rho e^{-i2\eta^+}-2\rho e^{-i\eta^+}\sum_{n=1}^N\frac{c_n}{k_n}
\psi_1(x,k_n)e^{i\lambda_n\zeta_nx}.
\end{equation}
Here
$$
\lambda_n=i\rho\sin\beta_n, \quad
\zeta_n=\rho\cos\beta_n,
$$
$$
\nu_n=\rho^2\sin(2\beta_n), \quad v_n=\rho^2(1+2\sin^2\beta_n),
$$
\begin{equation}
c_n(t)=c_n(0)e^{i2\lambda_n\zeta_n(2\lambda_n^2-\rho^2)t}=c_n(0)e^{\nu_nv_n t},
\end{equation}
and
\begin{equation}\label{eta}
\eta^+(x)=\frac12\int_x^\infty(\rho^2-|u|^2)dx.
\end{equation}
Symmetric relations such as
\begin{equation}
\tilde{\psi}(x,\rho^2k^{-1})=i\rho^{-1} k\sigma_3\psi(x,k),
\end{equation}
found in Ref.\cite{Chen04} are still valid. At $k=k_n$, we have
\begin{equation}
\tilde{\psi}(x,\bar{k}_n)=i\rho^{-1}k_n\sigma_3\psi(x,k_n),
\end{equation}
then, at $k=\bar{k}_m$, the first component of Eq.(\ref{IST}) is
\begin{equation}\label{IST3}
ie^{i\beta_m}\psi_1(x,k_m)= e^{-i\eta^+}e^{i\lambda_m\zeta_m x}
+2\sum_{n=1}^N\frac{k_nc_n\psi_1(x,k_n)}{\bar{k}_m^2-k_n^2}e^{i(\lambda_m\zeta_m+\lambda_n \zeta_n)x}.
\end{equation}
It can be shown that for $k_n$ located on the $\rho$-circle,
\begin{equation}
c_n=ie^{i\beta_n}\times(\mbox{a real number}).
\end{equation}
We can set
\begin{equation}
c_n(0)=i\chi_n\rho\sin(2\beta_n)e^{i\beta_n}e^{\nu_nx_n}, \quad \chi_n=\pm1.
\end{equation}
We also define
\begin{equation}
\theta_n=\nu_n(x-x_n-v_nt).
\end{equation}
In principle, one can find $\psi_1(x,k_m)$ by solving linear
equations
Eq.(\ref{IST3}) and then get a multi-soliton solution with Eq.(\ref{eta}) and
Eq.(\ref{u}). We only consider one-soliton and two-soliton solutions in this letter.

For the case of $N=1$, we get
\begin{equation}
u_1=\rho e^{-i2\eta^+_1}\frac{A_1}{{D}_1},
\end{equation}
where
\begin{equation}
D_1=1-i\chi_1e^{i\beta_1}e^{-\theta_1},
\end{equation}
\begin{equation}
A_1=1-i\chi_1e^{-i3\beta_1}e^{-\theta_1},
\end{equation}
\begin{equation}\label{eta1}
\eta^+_1=\frac12\int_x^\infty(\rho^2-|u_1|^2)dx=i\ln\frac{D_1}{\bar{D}_1},
\end{equation}
and thus the one-soliton solution,
\begin{equation}
u_1=\rho\frac{A_1D_1}{\bar{D}_1^2}=u_1(\theta_1),
\end{equation}
which is identical to that obtained in the literature. It is a bright soliton
for $\chi_1=-1$ or a dark soliton for $\chi_1=1$. There is only one parameter,
$\beta_1$, characterizing the soliton which is usually called one-parameter
soliton\cite{Mjolhus97}.

For the case of $N=2$, we get
\begin{equation}
u_2=\rho e^{-i2\eta^+_2}\frac{A_2}{D_2},
\end{equation}
where
\begin{eqnarray}\nonumber
D_2&=& 1-i\chi_1e^{i\beta_1}e^{-\theta_1}-i\chi_2e^{i\beta_2}
e^{-\theta_2} \\
& &-\chi_1\chi_2\frac{\sin^2(\beta_1-\beta_2)}
{\sin^2(\beta_1+\beta_2)}e^{i(\beta_1+\beta_2)}e^{-\theta_1-\theta_2},
\end{eqnarray}
\begin{eqnarray}\nonumber
A_2
&=&1-i\chi_1e^{-i3\beta_1}e^{-\theta_1}-i\chi_2e^{-i3\beta_2}
e^{-\theta_2}\\
& &-\chi_1\chi_2\frac{\sin^2(\beta_1-\beta_2)}
{\sin^2(\beta_1+\beta_2)}e^{-i3(\beta_1+\beta_2)}e^{-\theta_1-\theta_2}.
\end{eqnarray}
In order to find $\eta_2^+$, we find
\begin{equation}
{\rm Re}( D_2)\frac{d[{\rm Im}(D_2)]}{dx}-{\rm Im}(D_2)\frac{d[{\rm
Re}(D_2)]}{dx}=\frac{\rho^2}4(|D_2|^2-|A_2|^2).
\end{equation}
With this relation, we get
\begin{equation}\label{eta2}
\eta^+_2=\frac12\int_x^\infty(\rho^2-|u_2|^2)dx=i\ln\frac{D_2}{\bar{D}_2},
\end{equation}
and the two-soliton solution,
\begin{equation}
u_2=\rho\frac{A_2D_2}{\bar{D}_2^2}.
\end{equation}
Assume $\beta_2>\beta_1$. At times long before collision ($t\to-\infty$), in the
vicinity of
$\theta_1\approx 0$, $\theta_2\to\infty$, $u_2\approx u_1(\theta_1)$, while in the
vicinity
of $\theta_2\approx 0$, $\theta_1\to-\infty$, $u_2\approx u_1(\theta_2+\Delta)$, that
is,
\begin{equation}
u_2\approx u_1(\theta_1)+u_1(\theta_2+\Delta),
\end{equation}
where
\begin{equation}
\Delta=2\ln\left|\frac{\sin(\beta_1+\beta_2)}{\sin(\beta_1-\beta_2)}\right|>0.
\end{equation}
$u_2$ consists of two well separated solitons, moving to the
positive direction of the $x$-axis, with the slower soliton of $\beta_1$ moving
on the front.

At times long after collision ($t\to\infty$),
in the vicinity of $\theta_1\approx 0$,
$\theta_2\to -\infty$, $u_2\approx u_1(\theta_1+\Delta)$, while in the vicinity of
$\theta_2\approx 0$, $\theta_1\to \infty$, $u_2\approx u_1(\theta_2)$, that is,
\begin{equation}
u_2\approx u_1(\theta_1+\Delta)+u_1(\theta_2).
\end{equation}
Now $u_2$ consists of two well separated solitons, with the faster soliton of $\beta_2$
moving ahead.
These asymptotic solutions show that the faster soliton of $\beta_2$ catches up the
soliton of $\beta_1$, collides with the latter and leaves it behind, gets a forward shift $\Delta
x_2=\Delta/\nu_2$ in position while the soliton of $\beta_1$ gets a backward
shift $\Delta x_1=-\Delta/\nu_1$ in position. These shifts in position are independent
of $\chi_1$ and $\chi_2$, i.e., the bright or dark characters of the two
participating solitons. They are in agreement with those in Ref.\cite{Kawata79}
obtained by discussing the asymptotic behavior of the modulus of
an implicit two-soliton solution.

\begin{figure}[h]
\centering
\includegraphics[width=0.8\textwidth]{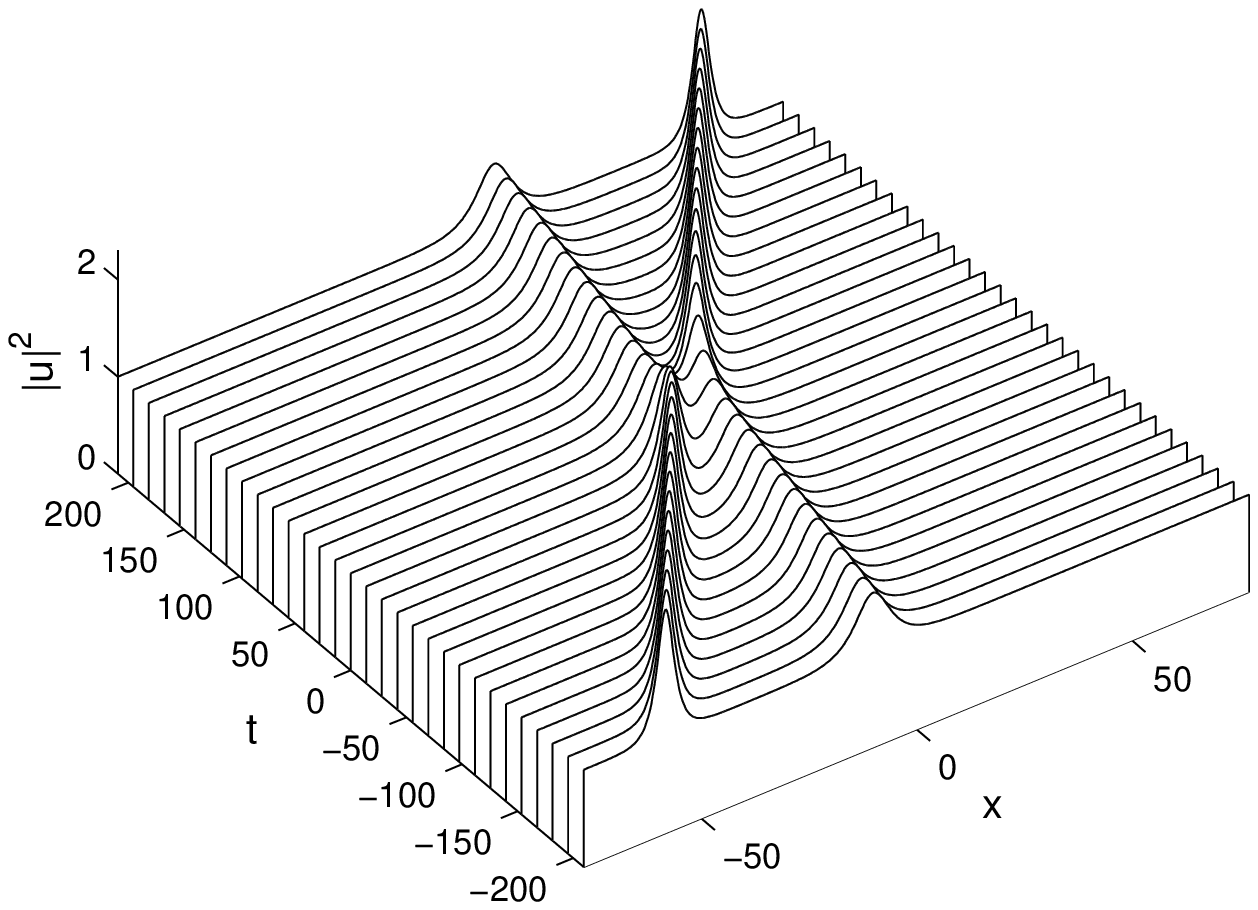}
\caption{\label{dnlsbb}Collision between two bright solitons, $\rho=1$,
$\beta_1=\pi/12$, $\beta_2=\pi/24$. Variables in the figure are dimensionless.}
\end{figure}

\begin{figure}[h]
\centering
\includegraphics[width=0.8\textwidth]{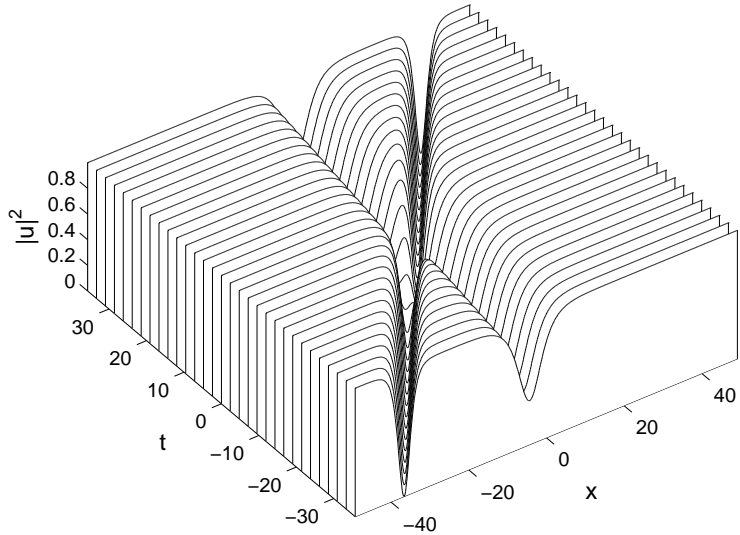}
\caption{\label{dnlsdd}Collision between two dark solitons, $\rho=1$, $\beta_1=\pi/15$,
$\beta_2=\pi/6$. Variables in the figure are dimensionless.}
\end{figure}

\begin{figure}[h]
\centering
\includegraphics[width=0.8\textwidth]{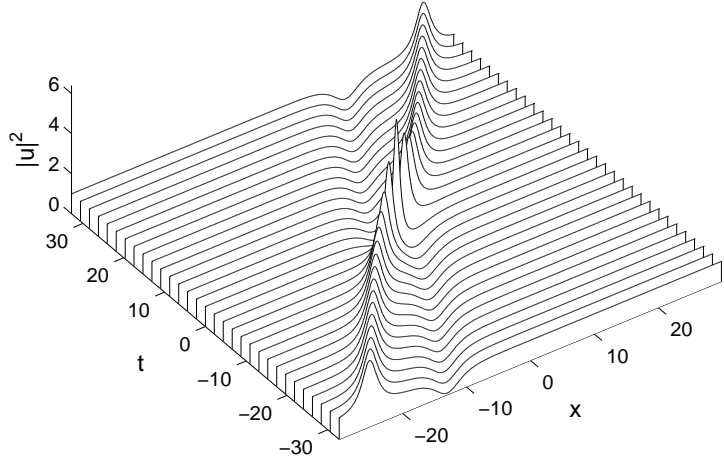}
\caption{\label{dnlsbd}A bright soliton of $\beta_2=2\pi/15$ chases a dark
soliton of $\beta_1=7\pi/60$, $\rho=1$. Variables in the figure are dimensionless.}
\end{figure}

\begin{figure}[h]
\centering
\includegraphics[width=0.8\textwidth]{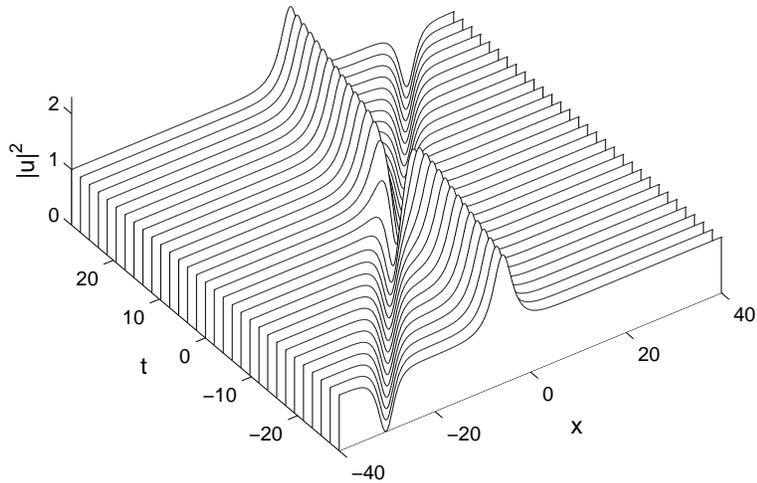}
\caption{\label{dnlsdb}A dark soliton of $\beta_2=\pi/6$ chases a bright
soliton of $\beta_1=\pi/12$, $\rho=1$. Variables in the figure are dimensionless.}
\end{figure}

Collisions between two bright solitons(Fig.\ref{dnlsbb}), two dark solitons
(Fig.\ref{dnlsdd}),
and one bright soliton and one dark soliton(Fig.\ref{dnlsbd} and Fig.\ref{dnlsdb})
are graphically shown, where shifts of soliton positions due to collisions are vividly
seen. We choose $x_1=\Delta/(2\nu_1)$ and $x_2=\Delta/(2\nu_2)$
in these figures so that the two solitons completely overlap at $(x,t)=(0,0)$.

In summary, we find an explicit two-soliton solution for the DNLS equation with NVBC. Shifts
of soliton positions due to collisions between solitons are analytically obtained.
Details of collisions between two bright solitons, two dark solitons and one bright
soliton and one dark soliton are graphically shown. It is interesting to note that
these shifts only depend on parameters of the
participating solitons, irrespective of their bright or dark characters.
Observing Eq.(\ref{eta1}) and
Eq.(\ref{eta2}), one can find that the relations between $\eta$ and $D$ for
one-soliton and two-soliton are the same. The relation can possibly be extended
to multi-soliton case and be helpful to find an explicit multi-soliton solution.

\newpage
\end{document}